\documentclass[conference]{IEEEtran}
\IEEEoverridecommandlockouts

\usepackage{cite}
\usepackage{amsmath,amssymb,amsfonts}
\usepackage{graphicx}
\usepackage{textcomp}
\usepackage{xcolor}
\usepackage{array}
\usepackage{url}
\usepackage{comment}
\usepackage{amsmath}
\usepackage{amsfonts}
\usepackage{amssymb}
\usepackage{siunitx}
\usepackage{booktabs}
\usepackage[linesnumbered,ruled,vlined]{algorithm2e}
\usepackage[noend]{algpseudocode}
\usepackage{mathtools}

\usepackage[symbol]{footmisc}
\def\BibTeX{{\rm B\kern-.05em{\sc i\kern-.025em b}\kern-.08em
    T\kern-.1667em\lower.7ex\hbox{E}\kern-.125emX}}
    
\begin{document}

\title{qBSA: Logic Design of a 32-bit \\Block-Skewed RSFQ Arithmetic Logic Unit
\thanks{This work was supported by the Office of the Director of National Intelligence (ODNI), the Intelligence Advanced Research Projects Activity (IARPA), via the U.S. Army Research Office Grant W911NF-17-1-0120.}
\thanks{Accepted in IEEE ISEC 2019.}}

\author{\IEEEauthorblockN{Souvik Kundu, Gourav Datta, Peter A.~Beerel, Massoud Pedram }
\IEEEauthorblockA{\textit{Ming Hsieh Department of Electrical and Computer Engineering} \\
\textit{University of Southern California}\\
Los Angeles, California 90089, USA \\
\{souvikku, gdatta, pabeerel, pedram\}@usc.edu}
}

\maketitle

\begin{abstract}
Single flux quantum (SFQ) circuits are an attractive beyond-CMOS technology because they promise two orders of magnitude lower power at clock frequencies exceeding 25 GHz. However, every SFQ gate is clocked creating very deep gate-level pipelines that are difficult to keep full, particularly for sequences that include data-dependent operations. This paper proposes to increase the throughput of SFQ pipelines by re-designing the datapath to accept and operate on least-significant bits (LSBs) clock cycles earlier than more significant bits. This {\em skewed} datapath approach reduces the latency of the LSB side which can be feedback earlier for use in subsequent data-dependent operations increasing their throughput. 
In particular, we propose to group the bits into 4-bit blocks that are operated on concurrently and create block-skewed datapath units for 32-bit operation. 
This skewed approach allows a subsequent data-dependent operation to start evaluating as soon as the first 4-bit block completes.
Using this general approach, we develop a block-skewed MIPS-compatible 32-bit ALU. Our gate-level Verilog 
design improves the throughput of 32-bit data dependent operations by 2x and 1.5x compared to previously proposed 4-bit bit-slice and 32-bit Ladner-Fischer ALUs respectively. 
We have quantified the benefit 
of this design on instructions per cycle (IPC) 
for various RISC-V benchmarks assuming a range 
of non-ALU operation latencies from one to ten cycles. 
Averaging across benchmarks, our experimental results 
show that compared to the 32-bit Ladner-Fischer our proposed 
architecture provides a range of IPC improvements 
between 1.37x assuming one-cycle non-ALU latency to 
1.2x assuming ten-cycle non-ALU latency.
Moreover, our average IPC improvements compared to a 
32-bit ALU based on the 4-bit bit-slice range from 2.93x to 4x.
\end{abstract}


\section{Introduction and Motivation}\label{intro}
The ever-increasing computational requirements of high performance computing (HPC) has leveraged the scaling of contemporary technologies for decades,
now reaching the atomic level.
However, the power density of silicon nano-electronics limits their applicability to future exascale computing \cite{esmaeilzadeh2011dark,borkar2011future}, motivating the research for alternate technologies. Evolved from rapid SFQ (RSFQ) \cite{likharev1991} technology, superconductive circuits that 
promise ultra-low switching energy of $10^{-19}$J \cite{volkmann2013experimental} and clock frequencies exceeding 25GHz \cite{tang20154} have become a promising beyond-CMOS technology.

Various 8-bit SFQ microprocessors have been developed in the last two decades, including a bit-serial microprocessor with
eight 1-bit serial ALU blocks (FLUX-1) \cite{dorojevets2001flux}, 
a bit-serial CORE1 processor \cite{fujimaki2008bit}, and
a bit-serial SCRAM2 asynchronous microprocessor \cite{nobumori2007design}. 
More specifically, the arithmetic logic unit (ALU), a critical part of a microprocessor, 
has gained significant research importance in RSFQ \cite{filippov20118}, \cite{filippov201220}, \cite{tang20164}, \cite{dorojevets20138}. 
Recently Tang et. al. have proposed a 16-bit bit-sliced ALU \cite{tang2018logic} because earlier proposed serial \cite{ando201580} and 2-/4-/8-bit bit-sliced \cite{tang20164} ALUs compute at a slower rate for 32-/64-bit processors. 
As we increase the ALU bit-width, its gate-level pipelined nature, forces an increase in latency and efficiently utilizing this deep pipelined architecture becomes more difficult. 

To improve pipeline utilization we propose a block-skewed ALU architecture, called qBSA, inspired by the use of skewed datapaths in asynchronous CMOS design \cite{manohar2001width}. Our proposed architecture uses eight 4-bit ALU blocks skewed in time, reduces the delay of the data feedback loop, and enables individual blocks to start computing a dependent operation as soon as its own output is ready. 
The choice of 4-bit blocks enables a balance between keeping the latency of the 32-bit adder relatively low while requiring fewer Josephson junctions (JJ) than needed for higher bit-width blocks.
We have simulated our results using the MIT LL 100$\mu$A/$\mu$m$^{2}$ SFQ5ee RSFQ cell library to demonstrate its functional correctness. We have also estimated 
its impact on the instructions per cycle (IPC) 
of a RISC-V processor. 

The reminder of this paper is arranged as follows. Section \ref{arch} describes the proposed architecture and explains its functionality. Section \ref{res} provides our simulation platform, results and performance analysis. Finally, the paper concludes in Section \ref{conc}.

\section{Proposed 32-bit Block-skewed architecture}\label{arch}
\begin{figure*}[t]
\centering
\includegraphics[width = 1\linewidth]{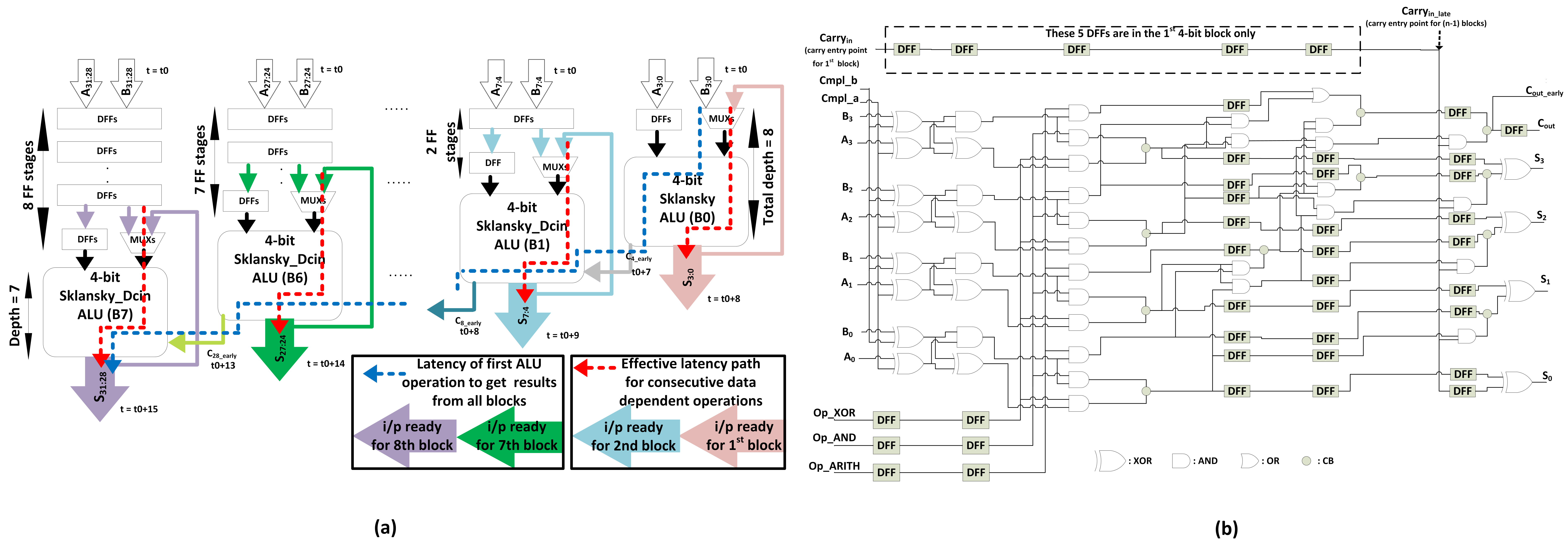}
\caption{(a) Proposed micro-architecture of our 32-bit block-skewed ALU with eight 4-bit Sklansky ALU blocks (b)  Gate level diagram of the proposed 4-bit Sklansky ALU block. Note that the first 4-bit ALU block is different from the other 7, because its carry input arrives at the same time as its A and B inputs. In contrast, the carry input for the 7 other blocks arrives 5 clock cycles after their A and B inputs.}
\label{fig-4b_alu_ckt}
\end{figure*}

In this section we describe the logic design of our 32-bit qBSA.
We divided the design into eight 4-bit blocks as shown in Fig. \ref{fig-4b_alu_ckt}(a). 
Due to its low latency, and simple carry look ahead circuit with only one feed forward signal ($c_{out}$) we adopted the Sklansky prefix-tree adder \cite{sklansky1960conditional} to design each 4-bit block, as illustrated in Fig. \ref{fig-4b_alu_ckt}(b).
Notice that the carry ($C_{in}$) is needed to compute the carry out ($C_{out\_early}$, $C_{out}$) and sum ($S_{n+3:n}$) only after five pipeline stages. 
We leverage this fact and start computing the sum and carry of more significant blocks before the $C_{out\_early}$ of the less significant blocks are evaluated.
It is to be noted that we use $C_{out\_early}$ to quickly feed the input carry of the next 4-bit ALU block and delay it by one stage to provide the final $C_{out}$. 
The feedback path from the output of each block back to its input (through a multiplexer) enables less significant blocks to start accepting and computing their next data-dependent inputs as soon as the previous corresponding output is ready, thereby avoids waiting for the entire 32-bit result. This staggers the computation start time for different blocks making the datapath {\em skewed} and better utilizes the gate-level pipelining nature of SFQ. In particular, this reduces the {\em initiation interval} ({\em II}) for back-to-back data-dependent 
operations, defined as the number of clock-cycle separation between the start of two consecutive data-dependent operations. 

Table \ref{tab:op_list} shows the operations supported by qBSA and their associated control signals. 
Table \ref{tab:lat_iii_vals} shows the latency and the {\em initiation interval} values of our proposed design.

\begin{table}[!ht]
    \caption{Supported ALU Instructions}
    \centering
    \begin{tabular}{|c|c|c|c|c|c|c|}
    \hline
         ALU & $Op\_$ & $Op\_$ & $Op\_$ & $Cmpl\_$ & $Cmpl\_$ & $C_{in}$\\ 
         Operations & ARITH & AND & XOR & a & b & \\
    \hline
         ADD & 1 & 0 & 1 & 0 & 0 & 0 \\
    \hline
         SUB & 1 & 0 & 1 & 0 & 1 & 1 \\
    \hline
         SLT & 1 & 0 & 1 & 0 & 1 & 1 \\
    \hline
         EQ & 0 & 0 & 1 & 0 & 1 & 1 \\
    \hline
         AND & 0 & 1 & 0 & 0 & 0 & 0 \\
    \hline
         OR & 0 & 1 & 1 & 0 & 0 & 0 \\
    \hline
         XOR & 0 & 0 & 1 & 0 & 0 & 0 \\
    \hline
         NOR & 0 & 1 & 0 & 1 & 1 & 0 \\
    \hline
    \end{tabular}
    \label{tab:op_list}
\end{table}

\begin{table}[!ht]
    \caption{Latency and Initiation Interval of our proposed Block-skewed ALU}
    \centering
    \begin{tabular}{|c|c|c|}
    \hline
        Parameter & Data Dependency & Pipeline stages   \\
    \hline
        Latency & N/A & 15 \\
    \hline
        Initiation Interval & Yes & 8 \\
    \hline
        Initiation Interval & No & 1 \\    
    \hline
    \end{tabular}
    \label{tab:lat_iii_vals}
\end{table}

\section{Results}\label{res}
We used Verilog models of a 100$\mu$A/$\mu$m$^2$ MIT LL SFQ5ee cell library to design and simulate qBSA in the Xilinx Vivado 2017.4 tool. Note that in our simulated waveforms a signal transition (high to low or vice versa) and no transition represent presence and absence of SFQ pulse, respectively.    
\subsection{Gate-level Simulation}\label{sim_res}
\begin{figure}[!t]
\centering
\includegraphics[width = 1\linewidth]{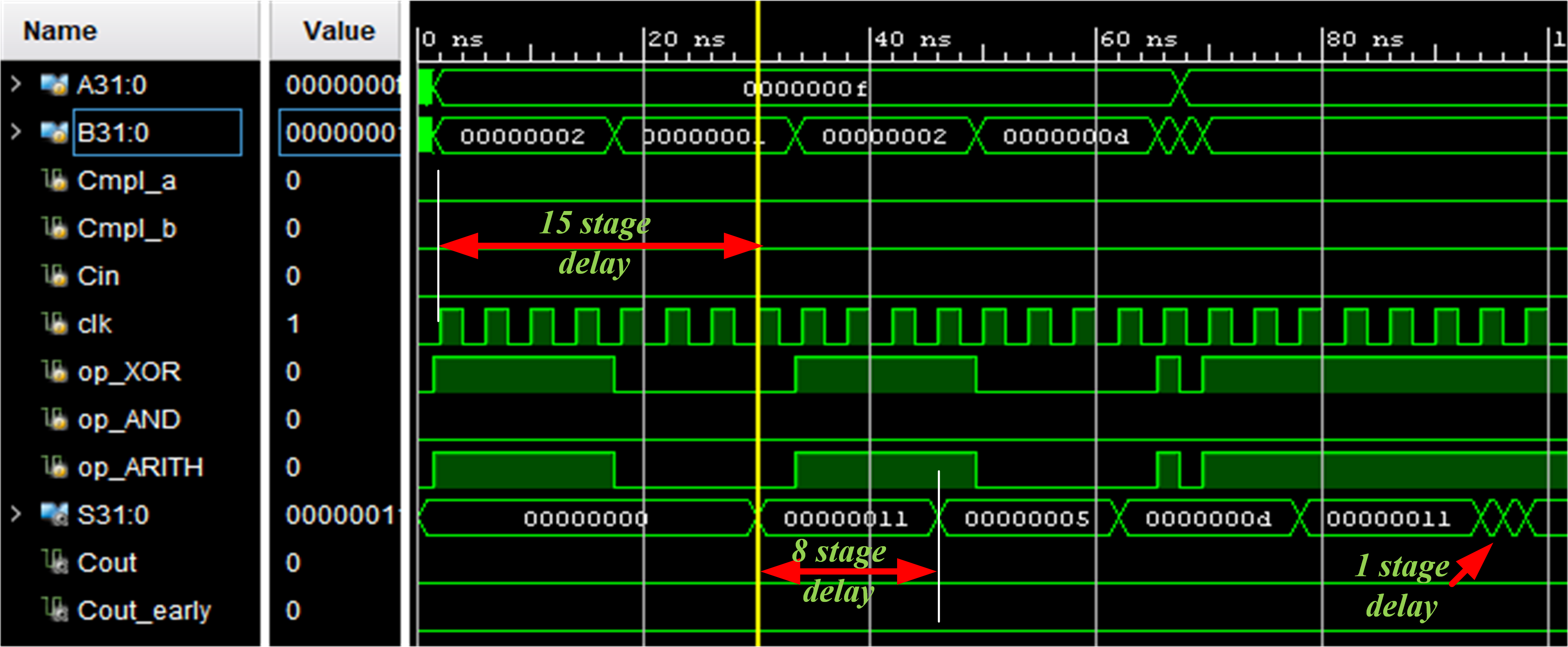}
\caption{Gate-level simulation of an ADD operation in our proposed qBSA.}
\label{fig-behav_sim}
\end{figure}

Fig. \ref{fig-behav_sim} shows a typical waveform generated through gate-level simulation of the proposed 32-bit ALU. Notice that after the first output is available, the skewed datapath of the qBSA enables back-to-back data-dependent outputs available after the pipeline depth of a 4-bit ALU block (8-clock stages) instead of the pipeline depth delay of the entire 32-bit ALU (15-clock stages).  Thus the {\em initiation interval} of our proposed qBSA is 1.5x and 2x faster compared to recently proposed 32-bit Ladner Fischer ALU (32LFA) \cite{tang2018logic} and 4-bit bit sliced ALU (4BSA) \cite{tang20154}, respectively.\footnote{For both the 32LFA and 4BSA ALUs we have added a 1-clock delay for the MUX-stage to their actual stage delays to perform 32-bit data-dependent operations, obtaining {\em II}s of 12 and 16, respectively.}  

\subsection{Performance Evaluation: Instruction Per Cycle}\label{IPC}

To quantify the benefit of our proposed design we estimated the impact on IPC for a set of benchmarks 
on a generic qBSA-based RISC-V processor with in order commitment (qBSP). We compared the obtained IPC to that of a 32LFA (32LFP) and 4BSA (4BSP) based processors. In particular, the IPC of a benchmark with total number 
of instructions $T_i$ and total NOPs needed to resolve dependencies $T_{NOP}$ is as follows: 
\begin{IEEEeqnarray}{c}\label{eq-IPC}
IPC = \frac{T_i}{(T_i+T_{NOP})}
\end{IEEEeqnarray}

We estimate the IPC using a script that reads benchmark files generated through Spike, a RISC-V sodor core instruction set architecture (ISA) simulator, analyzes the dependencies, and estimates the number of NOPs required \cite{spike1}. 
We assume all processor components are block-skewed and consume and generate inputs and outputs in block-skewed fashion. 
In particular, Equations \ref{eq-dep1} and \ref{eq-dep2} recursively defines the number of NOPs required before each instruction $i$ and its final position considering the added NOPs.
\begin{IEEEeqnarray}{c}\label{eq-dep1}
\begin{multlined}
NOP[i] = max(0, max_{m \in N_{S_i}}(L({S(i,m))} -\\ 
(pos[i-1] +1 - pos[I(i,m)]) \; )) \IEEEyesnumber
\end{multlined}
\end{IEEEeqnarray}
\begin{IEEEeqnarray}{c}\label{eq-dep2}
\begin{multlined}
pos[i] = pos[i-1] +NOP[i] + 1
\end{multlined}
\end{IEEEeqnarray}
Here, functions {\em S(i,m)} and {\em I(i,m)} provide the instruction type and original index of the instruction that creates the $m^{th} \in N_{S_{i}}$ source operand of the $i^{th}$ instruction. 
$L({S(i,m)})$ is the latency of the instruction which creates the $m^{th}$ source register of instruction $i$. 

\begin{figure}[!ht]
\centering
\includegraphics[width=1\linewidth]{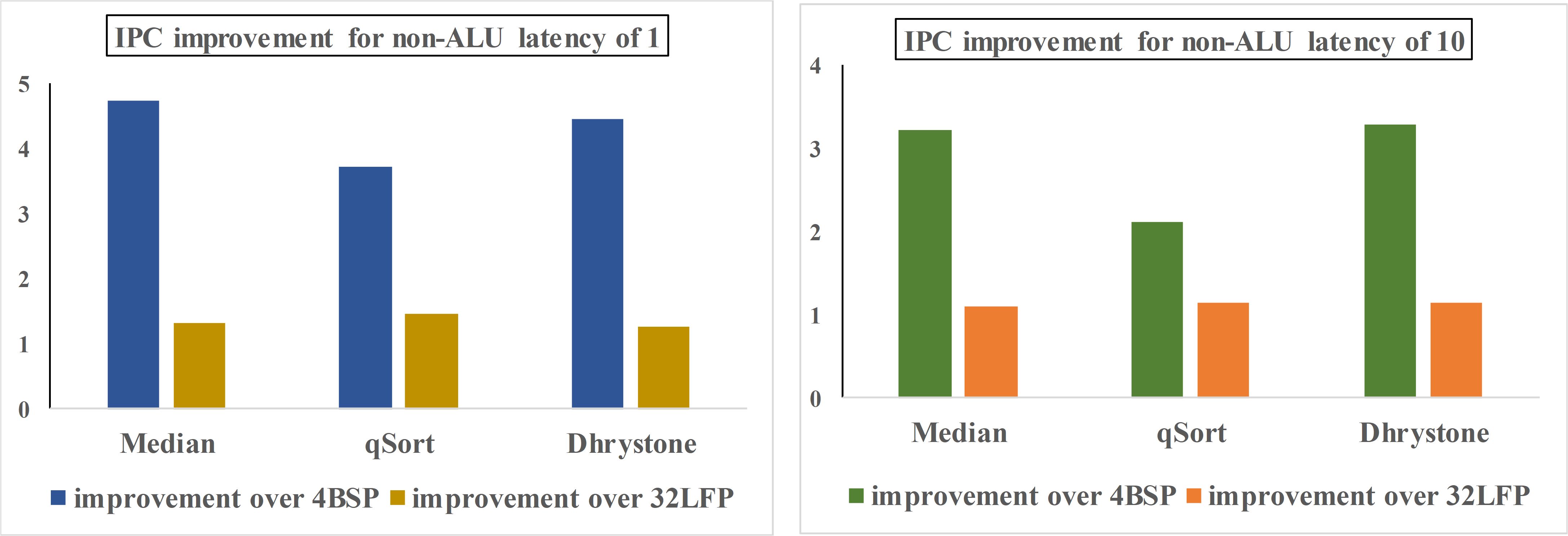}
\caption{IPC comparison of qBSP, 32LFP, and 4BSP for non-ALU latencies of 1 and 10 for three different benchmarks.}
\label{fig-IPC_graph}
\end{figure}

Our experiments explore a range of non-ALU data-dependent operation latencies $[1, 10]$ but in each individual experiment, for simplicity, we assume that all non-ALU operations have the same integral latency.
As two examples, Fig. \ref{fig-IPC_graph} shows the IPC improvement of qBSP over 32LFP and 4BSP with non-ALU latency assumptions 1 and 10. 

\section{Conclusions}\label{conc}

The gate-level pipelined nature of RSFQ makes keeping the pipelines 
full a difficult micro-architectural challenge, especially in the presence of data-dependent operations. This paper proposes a block-skewed ALU 
to reduce the average pipeline initiation interval 
and estimates its impact on an ideal RSFQ processor.
Averaging across multiple benchmarks with a simple dependency model, block-skewing improves 
IPC between 1.2x and 1.37x 
compared to a 32-bit Ladner Fischer ALU based processor and between 2.93x and 4x compared to a 4-bit bit-sliced ALU based processor. 
Our future work includes evaluating the benefits of block skewing on 
other processor components, the impact of different block sizes, 
and refinements of our model of instruction dependencies.    
\bibliographystyle{IEEEtran}
\bibliography{IEEEabrv,bibliography}

\end{document}